\documentclass[aps,prl,twocolumn,superscriptaddress,nofootinbib,floatfix,raggedbottom]{revtex4-2}
\usepackage{amsmath,amssymb,mathtools,amsthm}
\usepackage{bm}
\usepackage{microtype}
\usepackage{tikz}
\usetikzlibrary{decorations.markings,arrows.meta}
\usepackage[colorlinks=true,linkcolor=blue,citecolor=blue,urlcolor=blue]{hyperref}

\newcommand{\HH}{\mathcal H}
\newcommand{\LB}{\mathcal L}
\newcommand{\A}{\mathcal A}
\newcommand{\Ap}{\mathcal A'}
\newcommand{\At}{\widetilde{\mathcal A}}
\newcommand{\B}{\mathcal B}
\newcommand{\Z}{\mathcal Z}
\newcommand{\PA}{\mathbb P_{\A}}
\newcommand{\PAp}{\mathbb P_{\Ap}}
\newcommand{\PZ}{\mathbb P_{\Z}}
\newcommand{\Tr}{\operatorname{Tr}}
\newcommand{\Id}{\mathbb I}
\newcommand{\id}{\operatorname{id}}
\newcommand{\Mat}{\operatorname{Mat}}

\newcommand{\AlgSet}{\mathbb A}
\newcommand{\btau}{\boldsymbol\tau}
\newcommand{\Ud}{\mathrm U(d)}
\newcommand{\Rsp}{\mathcal R}
\newcommand{\PR}{\mathbb P_{\Rsp}}
\newcommand{\Nrm}{N(\A)}
\newcommand{\uLie}{\mathfrak u(d)}
\newcommand{\nLie}{\mathfrak n}
\newcommand{\mLie}{\mathfrak m}

\newcommand{\hol}{\mathrm{hol}}
\newcommand{\Hol}{\mathrm{Hol}}
\newcommand{\holLie}{\mathfrak{hol}}
\newcommand{\Uofb}{\mathcal U}

\DeclareMathOperator{\Ad}{Ad}
\DeclareMathOperator{\spann}{span}

\theoremstyle{plain}
\newtheorem{theorem}{Theorem}
\newtheorem{conjecture}{Conjecture}

\newcommand{\su}{\mathfrak{su}}
\begin{document}

\title{Universal holonomic control of algebras of observables}

\author{Paolo Zanardi}
\affiliation{Department of Physics and Astronomy, and Center for Quantum Information
Science and Technology, University of Southern California, Los Angeles, California
90089-0484, USA}
\affiliation{Department of Mathematics, University of Southern California,
Los Angeles, California 90089-2532, USA}

\date{\today}

\begin{abstract}
\tolerance=1000
\looseness=-1
We extend geometric parallel transport from states and degenerate eigenspaces
to observable algebras defining generalized quantum subsystems. At fixed block structure,
unitary invariance fixes a unique connection, recovering Berry--Wilczek--Zee
transport on bands. On positive-dimensional strata, its holonomy is the
special-unitary part of the normalizer: all inner logical gates and equal-type
sector permutations are attainable. Given a compatible gapped Hamiltonian and
access to the required paths, inner gates can be
implemented adiabatically by isospectral control of a band refinement.
\end{abstract}

\maketitle

\emph{Introduction.}---The identification of the parts of a quantum system
determines what can be controlled locally, where information is stored, and
which states are entangled. These parts need not coincide with the physical
constituents of a device. The algebraic approach to quantum subsystems
\cite{VQS,VQS_Zanardi2004,Subsystems} identifies them through the observables
that can be measured and manipulated---their \emph{observable algebras}.
General algebras define \emph{generalized quantum subsystems}: their direct-sum
sectors carry a classical label, while each sector has a tensor-product structure.
This operational viewpoint encompasses virtual tensor factors, noiseless
subsystems \cite{DFS,DFS_Lidar1998,DFS_Knill2000}, and the logical degrees of
freedom of operator error-correcting codes
\cite{KribsLaflammePoulin05,Kribs2006,BenyKempfKribs2007}; it also underlies
generalized notions of entanglement and subsystem structure
\cite{Subsystems_Barnum2004,Subsystems_Mestoudjian2025}.

This viewpoint also connects to \emph{quantum mereology}, which seeks
physically preferred subdivisions from the dynamics
\cite{Mereology_CarrollSingh2021}. In the operational approach of Zanardi,
Dallas, Andreadakis, and Lloyd \cite{Mereology_Zanardi2024}, an algebra and
its commuting partner specify generalized subsystems, and minimal short-time scrambling
selects divisions that best retain their informational identity. Families of
observable algebras thus provide a setting for studying both subsystem
stability and control.

Here we ask how observables should be carried when the algebra specifying
them itself changes. Consider a logical qubit whose embedding in a larger
register is varied around a closed loop. Returning its observable algebra
to the initial set need not return each logical observable to itself:
parallel transport can leave a nontrivial logical gate. To determine this
operation one needs a transport rule for the algebra and a characterization
of the transformations that closed paths can generate.

\begingroup
\tolerance=1000
\looseness=-1
Geometric phases provide a complementary perspective on controlled changes
of quantum structures. A state or degenerate eigenspace can return to its
initial configuration while its transported frame acquires a Berry phase
or a Wilczek--Zee (WZ) rotation
\cite{Berry,Simon1983,WZ,AharonovAnandan1987,Anandan1988}.
Holonomic quantum computation uses these transformations as quantum gates
\cite{HQC,HQC_Unanyan1999,HQC_PachosZanardiRasetti1999,HQC_PachosZanardi2001,HQCrev,HQC_Zhang2023,HQC_Liang2023}.
Their use for virtual subsystems was already envisaged in Ref.~\cite{VQS},
and Oreshkov established universal holonomic computation in encoded
subsystems~\cite{HQCsub_Oreshkov2009}.
\par
\endgroup

In this work we extend geometric parallel transport from quantum states and
degenerate subspaces to the observable algebras that define generalized quantum
subsystems.
Algebras of fixed block structure form a homogeneous configuration space, or
stratum, on which unitary invariance uniquely fixes the transport of their
internal frames. Berry phases and Wilczek--Zee rotations are recovered on
spectral bands. We determine the full holonomy of this connection and thereby
the logical operations generated by closed paths of subsystem embeddings.
Whenever the configuration space has positive dimension, closed
paths realize every inner logical operation within each sector, as well as
permutations of sectors with identical block structure. Given a compatible
gapped Hamiltonian and access to the
required control paths, the logical operations can be implemented adiabatically
through a refinement into spectral bands while keeping the energy spectrum
fixed. For virtual tensor factors, only the interaction component of a control
changes the factorization. Proofs and numerical checks are provided in Appendices~\ref{sec:setup}--\ref{sec:scope}.

\emph{Transport of observable algebras.}---Let $\A\subseteq\LB(\HH)$ be a
unital $*$-algebra on $\HH\cong\mathbb C^d$, with commutant $\Ap$ and center
$\Z=\A\cap\Ap$. The Artin--Wedderburn decomposition reads
\begin{equation}\label{eq:wedderburn}
\begin{split}
   &\HH=\textstyle\bigoplus_{J=1}^{Z}\mathbb C^{n_J}\otimes\mathbb C^{d_J},\\
   &\A=\textstyle\bigoplus_J \Id_{n_J}\otimes\Mat(\mathbb C,d_J),\ \
   \Ap=\textstyle\bigoplus_J \Mat(\mathbb C,n_J)\otimes\Id_{d_J},
\end{split}
\end{equation}
with $\Z=\bigoplus_J\mathbb C\Pi_J$; its minimal projections $\Pi_J$ label
superselection sectors. In sector $J$, $d_J$ is the matrix-block size and $n_J$
its multiplicity: $\mathbb C^{d_J}$ carries the logical subsystem and
$\mathbb C^{n_J}$ the multiplicity factor. The center describes the classical
sector label; $Z=1$ gives a factor algebra and an ordinary tensor-factor
subsystem. Band structures and projective measurements (all $d_J=1$ or all
$n_J=1$) and noiseless codes are further instances.

Algebras with the same block sizes and multiplicities have the same
\emph{type}, specified by the multiset $\btau=\{(n_J,d_J)\}$, and form a
compact unitary orbit, the \emph{type stratum}:
$\AlgSet(\btau)=\{U\A U^\dagger:U\in\Ud\}\cong\Ud/\Nrm$ with
$\Nrm=\{U:U\A U^\dagger=\A\}$ the normalizer.
This description follows from Artin--Wedderburn and the homogeneous geometry
of conditional-expectation orbits \cite{CEorbits,CEorbits_AS,CEorbits_AL}.
The finite family of type strata provides a common setting for transport of
spectral decompositions, encoded and noiseless subsystems, and virtual tensor
factors. The stratum has dimension
$d^2-\sum_J(n_J^2+d_J^2-1)$ (Appendix~\ref{sec:setup}).

We call $\AlgSet(\btau)$
a \emph{band stratum} if every sector has $n_J=1$ or $d_J=1$; the algebra is then
fixed by an orthogonal decomposition of $\HH$ whose summands are labelled by their
type --- a partial spectral decomposition.

Motion is generated by unitaries, $\B(t)=U(t)\A U(t)^\dagger$.
We identify a tangent vector $UX$ at $U$ with $X$, so the velocity is
represented by $\Omega=U^\dagger\dot U\in\uLie$, where $\uLie$ is the
space of anti-Hermitian matrices.
A control generator splits into a component that preserves $\A$ as a set
and a component that moves it. The former is \emph{gauge} with respect to the
embedding, although it may act nontrivially on logical observables.

The gauge directions are now explicit. They are
$\Omega\in\nLie=(\A+\Ap)\cap\uLie$, the Lie algebra of $\Nrm$. Writing $\Uofb(\A)$ for the unitary group of $\A$, the identity
component of $\Nrm$ is $\Nrm_0=\Uofb(\Ap)\Uofb(\A)=\{\bigoplus_JU_J\otimes W_J\}$,
while the group of connected components $\pi_0(\Nrm)=\Nrm/\Nrm_0$ permutes sectors of equal type. The remaining directions move
$\A$. Let $\PA,\PAp,\PZ$ be the Hilbert--Schmidt projections onto $\A,\Ap,\Z$,
which are also the trace-preserving conditional expectations. The \emph{leakage}
subspace is $\Rsp=(\A+\Ap)^\perp$; the terminology comes from population loss out
of an encoded subspace \cite{HQCleak,HQCleak_LiuYung2020}. With $\PR=\id-\PA-\PAp+\PZ$, we obtain the
orthogonal, $\Ad(\Nrm)$-invariant splitting
\begin{equation}\label{eq:split}
   \uLie=\nLie\oplus\mLie,\qquad \mLie=\Rsp\cap\uLie ,
\end{equation}
where $\mLie\perp\Id$ (hence traceless) and $\dim\mLie=\dim\AlgSet(\btau)$: indeed
the map $A\mapsto\frac{d}{ds}|_0e^{sA}\A e^{-sA}$ identifies $\mLie$ with
$T_\A\AlgSet(\btau)$. The horizontal component $\PR\Omega$ alone moves $\A$; the gauge component
acts within the logical and multiplicity factors. Horizontal directions also determine
the leading leakage from the initial algebra, an averaged squared commutator
related to out-of-time-order correlators \cite{OTOC,OTOC_Maldacena2016} and algebraic
scrambling \cite{AOTOC,AOTOC_Andreadakis2023,AOTOC_Zanardi2024,Scrambling2_ZanardiLewis2026}
(Appendix~\ref{sec:setup}). The orbit and reductive splitting \cite{Nomizu1954},
including a prior uniqueness result, arise in conditional-expectation geometry
\cite{CEorbits,CEorbits_AS,CEorbits_AL}. Here we give a direct finite-dimensional
equivariance proof and determine the full holonomy across algebra types
(Appendices~\ref{sec:unique}--\ref{sec:curv}).

To determine the operation implemented by a loop, we need a rule for carrying the
internal frame of $\A$. Geometrically this is a principal connection on
$\Nrm\hookrightarrow\Ud\xrightarrow{\pi}\AlgSet(\btau)$,
$\pi(U)=U\A U^\dagger$. Equation \eqref{eq:split} defines the $\nLie$-valued gauge potential
\begin{equation}\label{eq:omega}
   \omega_U(U\Xi)=(\id-\PR)\,\Xi ,
\end{equation}
which extracts the gauge component. We call \eqref{eq:omega} the
\emph{canonical connection} associated with the orthogonal splitting
\eqref{eq:split}. A lift is horizontal when
$\omega_U(\dot U)=0$, equivalently $U^\dagger\dot U\in\mLie$.
Unitary invariance uniquely determines this connection: on every type stratum,
\eqref{eq:omega} is the \emph{only}
$\Ud$-invariant principal connection on $\Ud\to\AlgSet(\btau)$ (Appendix~\ref{sec:unique}).
This follows from Wang's classification of invariant connections
\cite{Wang1958,KobayashiNomizu}: the only $\Nrm$-equivariant map $\mLie\to\nLie$ is zero.
Thus the geometric transport used here is fixed by symmetry, and the agreement with
Berry--WZ transport below follows from uniqueness.

The construction is explicit. Given a smooth lift $V(t)$ of a curve of algebras
with $V(0)=U_0$, write $\Omega=V^\dagger\dot V$ and solve
$\dot h=-(\id-\PR)\Omega h$, $h(0)=\Id$. Then $h(t)\in\Nrm_0$ and
$\widetilde U=Vh$ is horizontal. The correction removes the
phase and all rotations that leave $\A$ fixed, including the inter-sector phases
and the independent rotations in $\A$ and $\Ap$. It is therefore the non-Abelian
analogue of Berry's removal of the dynamical phase.

For a loop $\gamma$, the holonomy
$\hol(\gamma)=U_0^\dagger\widetilde U(1)\in\Nrm$ is the operation implemented by
horizontal transport.
For a fixed initial frame, it is independent of the chosen lift; changing the initial frame conjugates it in
$\Nrm$, so its conjugacy class is intrinsic. In $\Nrm_0$ one may write
$\hol=(\oplus_JU_J\otimes\Id)(\oplus_J\Id\otimes W_J)$: the second factor is a
logical gate and the first rotates the multiplicity factors. A holonomy outside $\Nrm_0$
permutes sectors and is therefore outer (Fig.~\ref{fig:bundle}).

\begin{figure}[t]
\centering
\begin{tikzpicture}[scale=0.72,>=Stealth,line width=0.6pt]
\draw[gray,thick] (0,-0.05) -- (0,2.25);
\node[gray,anchor=west] at (0.10,2.20) {\scriptsize $\Nrm$};
\draw[red!75!black,thick,
      decoration={markings,mark=at position 0.55 with {\arrow{Stealth}}},
      postaction={decorate}]
  (0,0.80) .. controls (-1.35,0.92) and (-2.05,1.30) .. (-1.75,1.62)
          .. controls (-1.45,1.92) and (-0.60,1.90) .. (0,1.80);
\node[anchor=east] at (-1.95,1.28) {\scriptsize horizontal lift};
\fill (0,0.80) circle (1.3pt);
\fill (0,1.80) circle (1.3pt);
\draw[<->,thick] (0.17,0.80) -- (0.17,1.80);
\node[anchor=west] at (0.24,0.80) {\scriptsize $U_0$};
\node[anchor=west] at (0.24,1.80) {\scriptsize $U_0\,\hol(\gamma)$};
\draw[->,gray] (1.90,1.45) -- (1.90,0.55);
\node[gray,anchor=west] at (1.98,1.00) {\scriptsize $\pi$};
\draw[fill=blue!5] (-0.85,-0.60) ellipse [x radius=1.85, y radius=0.62];
\node[anchor=west] at (1.10,-1.02) {\scriptsize $\AlgSet(\btau)$};
\draw[red!75!black,thick,
      decoration={markings,mark=at position 0.40 with {\arrow{Stealth}}},
      postaction={decorate}]
  (-0.85,-0.60) ellipse [x radius=0.85, y radius=0.30];
\node[red!75!black,anchor=north] at (-1.15,-0.94) {\scriptsize $\gamma$};
\fill (0,-0.60) circle (1.3pt);
\node[anchor=west] at (0.08,-0.44) {\scriptsize $\A$};
\draw[gray,dashed,line width=0.4pt] (0,-0.05) -- (0,-0.60);
\end{tikzpicture}
\caption{A closed control loop $\gamma$ of algebras of type $\btau$: the lifted
frame fails to close by $\hol(\gamma)\in\Nrm$ --- a logical gate, a rotation
of the multiplicity factors, or a combination including a sector permutation.
The schematic was prepared with assistance from Anthropic Fable~5.1
under the author's direction and review.}
\label{fig:bundle}
\end{figure}

\emph{Intrinsic holonomy.}---We now classify the holonomy of
\eqref{eq:omega}. The restricted holonomy group $\Hol^0$ consists of holonomies
of contractible loops; its Lie algebra is $\holLie^0$
\cite{AmbroseSinger1953}. We write $X_{\nLie}$ for Hilbert--Schmidt
projection onto $\nLie$ and $\pi_1$ for the group of loop classes under
homotopies that fix the base point.

\begin{samepage}
\begin{theorem}\label{thm:hol}
(i) For horizontal $X,Y\in\mLie$, the curvature satisfies
$F_U(UX,UY)=-[X,Y]_{\nLie}\in\nLie$, independently of $U$.
(ii) Every holonomy has $\det\hol=1$.

(iii) The restricted holonomy algebra is exactly
$\holLie^0=\spann_{\mathbb R}\{[X,Y]_{\nLie}\}$, with no genericity assumption.

(iv) The transport is flat iff the stratum is zero-dimensional.

(v) The map $\pi_1(\AlgSet(\btau))\to\pi_0(\Nrm)$ assigning each loop class to the
component of its holonomy is onto: every permutation of equal-type
sectors is realized, and a loop is outer precisely when its class maps nontrivially.

(vi) On every positive-dimensional stratum, $\mLie$ generates
$\mathfrak{su}(d)$ and $\holLie^0=\nLie\cap\mathfrak{su}(d)$,
$\Hol=\Nrm\cap\mathrm{SU}(d)$ exactly.
\end{theorem}
\end{samepage}

By Theorem~\ref{thm:hol}(i), a small coordinate square of side $\varepsilon$ spanned by $X$ and $Y$ has
holonomy $\exp(\varepsilon^2[X,Y]_{\nLie})+O(\varepsilon^3)$. Theorem~\ref{thm:hol}(ii) removes a
continuous overall phase, but not the finite center of $\mathrm{SU}(d)$.
Theorem~\ref{thm:hol}(iv) shows that every positive-dimensional stratum has nontrivial holonomy.

The logical unitaries
$\Uofb(\A)=\{\oplus_J\Id_{n_J}\otimes W_J\}$ have determinant-one subgroup
$\mathcal{SU}(\A):=\Uofb(\A)\cap\mathrm{SU}(d)$.
Multiplying any $W\in\Uofb(\A)$ by a suitable global phase gives a
determinant-one representative with the same adjoint action. Thus conjugation by
determinant-one logical unitaries induces every inner automorphism of $\A$.
On every positive-dimensional stratum,
$\Hol=\Nrm\cap\mathrm{SU}(d)$ contains $\mathcal{SU}(\A)$, so every
inner logical gate is realized by a closed loop of algebras.
Theorem~\ref{thm:hol}(v) adds the discrete outer operations: permutations of
equal-type sectors are also realized by intrinsic closed loops.

\emph{The qubit stratum.}---The simplest positive-dimensional example is the
measurement algebra
$\A_{\mathbf n}=\spann\{\Id,\mathbf n\cdot\boldsymbol\sigma\}$, of type
$\{(1,1),(1,1)\}$. Because $\A_{\mathbf n}=\A_{-\mathbf n}$, its stratum is
$\mathbb{RP}^2$. The normalizer has two components: diagonal unitaries and their
products with $\sigma_x$. At $\A_{\mathbf z}$,
$\nLie=\spann\{i\Id,i\sigma_z\}$ and
$\mLie=\spann\{i\sigma_x,i\sigma_y\}$. Thus only the transverse part of a
control moves the axis, while horizontality transports both eigenstates by the Berry
rule. Write $\mathbf n(t)$ in spherical coordinates $\theta(t),\varphi(t)$ and use
$V=e^{-i\varphi\sigma_z/2}e^{-i\theta\sigma_y/2}$. The gauge part of
$V^\dagger\dot V$ is $-\frac i2\cos\theta\,\dot\varphi\,\sigma_z$, so the
horizontal correction is $h=\exp(\frac i2\sigma_z\int\cos\theta\,d\varphi)$. For a closed spherical path with
$\varphi(1)-\varphi(0)=2\pi$, $V(0)^\dagger V(1)=e^{-i\pi\sigma_z}=-\Id$, and
\begin{equation}\label{eq:qubithol}
   \hol=e^{-i\Omega_s\sigma_z/2},\qquad
   \Omega_s=\oint(1-\cos\theta)\,d\varphi .
\end{equation}
where $\Omega_s$ is the solid angle enclosed by the axis trajectory. The two outcomes therefore
see monopoles of charge $\mp\tfrac12$, with zero net phase by Theorem~\ref{thm:hol}(ii). Infinitesimally,
$F(i\sigma_x/2,i\sigma_y/2)=i\sigma_z/2$, while globally
$\Hol^0=\{e^{i\alpha\sigma_z}\}$. These continuous holonomies give opposite
phases to the eigenstates but act trivially on the commutative measurement
algebra. The noncontractible class of $\pi_1(\mathbb{RP}^2)=\mathbb Z_2$
instead exchanges its two minimal projections and is outer by
Theorem~\ref{thm:hol}(v). Indeed,
$U(t)=e^{-i\pi t\sigma_x/2}$ sends $\mathbf z$ to $-\mathbf z$ and hence returns
$\A_{\mathbf z}$ to itself. This is a loop of \emph{algebras}, not states; it is
already horizontal, with $\hol=-i\sigma_x$, and it swaps the two outcomes. Thus
$\Hol=\Nrm\cap\mathrm{SU}(2)=\{e^{i\alpha\sigma_z}\}
\sqcup\{-i\sigma_xe^{i\alpha\sigma_z}\}$.

For a physical duration $T$, the transverse Hamiltonian
$K=\hbar\pi\sigma_x/(2T)$ generates this horizontal exchange. As a gate on the
full qubit, this pulse is dynamical; its holonomy interpretation refers to the
closed loop of measurement algebras. Given access to couplings between
corresponding basis states, the same construction exchanges equal-type sectors.

\emph{Stabilizer codes.}---Let $S$ be a rank-$r$ Pauli stabilizer on $n$
qubits, $1\le r\le n$. Its $q=2^r$ syndrome sectors $\HH_x$ have dimension
$m=2^{n-r}$ and encode $n-r$ logical qubits each. The syndrome algebra
$\A_S=\spann_{\mathbb C}S$ and its commutant
$\A_S'=\bigoplus_x\LB(\HH_x)$ share the same canonical transport.
Theorem~\ref{thm:hol} gives
\begin{equation}\label{eq:stabilizer-hol}
  \Hol=\bigl(\mathrm U(m)^q\rtimes S_q\bigr)\cap\mathrm{SU}(2^n).
\end{equation}
Contractible loops realize $\Hol^0=S(\mathrm U(m)^q)$: independent
logical unitaries in each syndrome sector, subject to
$\prod_x\det V_x=1$. These unitaries commute with the syndrome algebra
$\A_S$ but can rotate logical observables in $\A_S'$. The component group
is $S_q$: every syndrome permutation occurs.

\emph{Tensor-product structures.}---A virtual tensor factor is specified by
$\btau=\{(d_1,d_2)\}$, with $d_1,d_2\ge2$, $d=d_1d_2$, and
$\A=\Id_{d_1}\otimes\Mat(\mathbb C,d_2)$ \cite{VQS,VQS_Zanardi2004,Scrambling2}. There is one
Wedderburn sector, so $\Nrm=\{V\otimes W\}$ is connected. (When $d_1=d_2$,
\textsc{swap} maps $\A$ to $\Ap$, not to itself.) The local controls
$ia\otimes\Id$ and $i\Id\otimes b$, with $a,b$ Hermitian, are gauge, while
$\mLie=\spann_{\mathbb R}\{i\,a\otimes b:\Tr a=\Tr b=0\}$ is the space of
horizontal interaction controls. Thus a general control moves the factorization iff
its interaction component is nonzero; the local part only changes the two internal
frames.

Theorem~\ref{thm:hol}(i) gives the curvature. For interaction controls
$i a\otimes b$ and $i c\otimes e$, the Hilbert--Schmidt overlap
$\Tr(be)$ on the logical factor weights the multiplicity rotation generated by $[a,c]$;
the overlap $\Tr(ac)$ on the multiplicity factor weights the logical rotation generated by $[b,e]$.
If both overlaps vanish, the curvature is zero. This also forces the full
commutator to vanish for every such pair only for two qubits (Appendix~\ref{sec:factor}).
For two qubits,
$\mLie=\spann_{\mathbb R}\{i\sigma_k\otimes\sigma_l\}$, and the small closed loop generated by $i\sigma_x\otimes\sigma_x$ and
$i\sigma_x\otimes\sigma_y$ implements
$\exp(-2i\varepsilon^2\Id\otimes\sigma_z)+O(\varepsilon^3)$.
\footnote{Only for $d_1=d_2=2$ do the commutators $[\mLie,\mLie]$ lie in
$\nLie$. For two qubits, this inclusion follows from the Pauli anticommutation
relations. In the magic basis, the stabilizer in $\mathrm{SU}(4)$ is
$\mathrm{SO}(4)\sqcup i\,\mathrm{SO}(4)$, giving the global symmetric stratum
$(\mathrm{SU}(4)/\mathrm{SO}(4))/\mathbb Z_2$ \cite{Helgason,Besse1987}
(Appendix~\ref{sec:factor}).} For every $(d_1,d_2)$, write
$k=\gcd(d_1,d_2)$:

\begin{samepage}
\begin{theorem}\label{thm:factor}
For $\btau=\{(d_1,d_2)\}$, $d_1,d_2\ge2$, the restricted holonomy group is
$\Hol^0=\mathrm{SU}(d_1)\otimes\mathrm{SU}(d_2)$, and
\begin{equation}\label{eq:Hol22}
   \Hol=\Nrm\cap\mathrm{SU}(d)
   =\bigsqcup\nolimits_{j=0}^{k-1}e^{2\pi ij/d}\,\Hol^0 ,
\end{equation}
so that $\Hol/\Hol^0\cong\pi_1(\AlgSet(\btau))\cong\mathbb Z_k$.
\end{theorem}
\end{samepage}

This specializes Theorem~\ref{thm:hol}(vi) to tensor factors; for two qubits,
$k=2$ and $\Hol^0\cong\mathrm{SO}(4)$. Contractible loops implement arbitrary
independent special-unitary rotations of the logical and multiplicity factors.
Every holonomy has the form $z_jh_0$, with $z_j=e^{2\pi ij/d}\Id$ and
$h_0\in\Hol^0$. Its action on observables is determined by $h_0$, while $j$
labels its connected component. The label is invariant under homotopy, but $h_0$ generally depends on the loop's shape.
These additional labels occur when $k>1$, are absent on band strata, and can be
generated by a controlled-phase loop (Appendix~\ref{sec:factor}).

\emph{Band strata and universal control.}---We now use spectral bands to
implement logical operations adiabatically.
On band strata, $\nLie$ consists of block-diagonal anti-Hermitian matrices, while $\mLie$
consists of the block-off-diagonal ones. For a moving frame
$|\psi_a(t)\rangle=\widetilde U(t)|a\rangle$, horizontality is
$\langle\psi_a|\dot\psi_b\rangle=0$ whenever $a,b$ belong to the same band.
This is precisely the WZ parallel-transport condition \cite{WZ}, applied to all
bands at once. Thus \eqref{eq:omega} gives the direct sum of the WZ connections.
For globally labelled bands, transport takes place on the \emph{ordered~band~flag}:
the space of orthogonal decompositions of $\HH$ into subspaces of prescribed
dimensions, each carrying a fixed eigenvalue label \cite{Spectral1993}. A closed
loop returns each labelled subspace to itself and gives a tuple of band holonomies.
A loop in the algebra stratum may instead permute equal-type
bands, correspondingly permuting the tuple. The Berry phase is the case of a
one-dimensional band. A closed control path in a gapped Hamiltonian family
with fixed band dimensions gives a loop of band structures. Its adiabatic
holonomy is that of \eqref{eq:omega} along this loop (Appendix~\ref{sec:band})
\cite{NarasimhanRamanan61}. No
adiabatic identification is claimed away from band strata.

For the control construction, consider isospectral families in which the
eigenbasis varies while the spectrum remains fixed:
\begin{equation}\label{eq:family}
   H(\lambda)=U(\lambda)H_0U(\lambda)^\dagger,\qquad U:M\to\Ud,\ \dim M\ge2 .
\end{equation}
To probe an algebra $\A$ of type $\btau$, assume a compatible reference Hamiltonian in its commutant,
$H_0=\bigoplus_JH_J\otimes\Id_{d_J}\in\Ap$, with each $H_J$ nondegenerate and
with disjoint spectra between sectors. Then $[H_0,\A]=0$, and sector $J$ is split
into $n_J$ bands of dimension $d_J$. Each band carries one copy of the logical
subsystem. Because the family \eqref{eq:family} is isospectral, all gaps remain fixed.
The transported object is the \emph{band algebra}
$\At=\{H_0\}''\vee\A=\bigoplus_{(J,a)}\Mat(\mathbb C,d_J)$, which is of band-stratum
type; $\vee$ denotes the algebra generated by its arguments. This refinement
allows independent operations on the $n_J$ copies of each logical factor, whereas
$\A$ requires the same operation on every copy.

Let $P_\mu$ project onto band $\mu=(J,a)$ of $H_0$. The anti-Hermitian
matrix-valued one-form $\theta=U^\dagger dU$ describes the change of the eigenframe
over $M$; $d$ denotes exterior differentiation. Its block
$\theta_{\mu\nu}=P_\mu\theta P_\nu$ maps reference band $\nu$ to $\mu$.
The diagonal block $A^{(\mu)}=\theta_{\mu\mu}$ is the WZ connection within band
$\mu$. The curvature of this connection,
$F^{(\mu)}=dA^{(\mu)}+A^{(\mu)}\wedge A^{(\mu)}$, governs the leading
holonomy around infinitesimal loops. Here $\wedge$ combines the exterior product
of forms with matrix multiplication. The identity $d\theta+\theta\wedge\theta=0$
gives $F^{(\mu)}=-\sum_{\nu\ne\mu}\theta_{\mu\nu}\wedge\theta_{\nu\mu}$:
the curvature within a band is determined by pairs of off-diagonal blocks
coupling its frame to the other bands.

For labelled bands, the joint WZ holonomy is block diagonal.
Applying Theorem~\ref{thm:hol}(i),(ii) gives
\begin{equation}\label{eq:sumrule}
   \prod_{(J,a)}\det h^{(J,a)}=1 ,
   \qquad
   \sum_{(J,a)}\Tr F^{(J,a)}=0
\end{equation}
The determinant identity holds for every closed loop in the control manifold,
including noncontractible ones; individual band determinants need not equal one. On the full ordered band
flag, this collective determinant constraint is the only restriction on the
joint holonomy:

\begin{theorem}[Holonomy of isospectral families]\label{thm:WZ}
For every family \eqref{eq:family} with at least two bands the joint WZ transport
preserves each band, and its holonomy group is contained in
\begin{equation}\label{eq:WZgroup}
   S\bigl(\textstyle\prod_J\mathrm U(d_J)^{n_J}\bigr)
   =\bigl\{(g_{J,a}):\prod_{J,a}\det g_{J,a}=1\bigr\}.
\end{equation}
This group has dimension $\sum_Jn_Jd_J^2-1$ and is exactly the holonomy group of
the canonical connection \eqref{eq:omega} on the ordered band flag
$\Ud/\prod_{J,a}\mathrm U(d_J)$ (Appendix~\ref{sec:isospectral}).
\end{theorem}

For the canonical connection on the ordered band flag, independent
$\mathrm{SU}(d_J)$ rotations are attainable in every band, while the band
determinants range over a torus of dimension $\sum_J n_J-1$. For a specified control
family, the holonomy group may be a proper subgroup of the group in
\eqref{eq:WZgroup}; a constant family has trivial holonomy. Whether equality holds
for generic families remains conjectural (Appendix~\ref{sec:isospectral}).

The original adiabatic construction of holonomic quantum computation acts
within a degenerate eigenspace
\cite{HQC,HQC_Unanyan1999,HQC_PachosZanardiRasetti1999,HQC_PachosZanardi2001},
with proposals across platforms
\cite{HQCimp,HQCimp_Falci2000,HQCimp_PachosChountasis2000,HQCimp_Duan2001,HQCimp_Faoro2003}.
Nonadiabatic geometric gates use cyclic state or subspace evolution
\cite{NHQC,NHQC_Sjoqvist2012,NHQC_Liu2019}, with experimental implementations
\cite{HQCexp,HQCexp_Feng2013,HQCexp_Zu2014,HQCexp_Sekiguchi2022,HQCexp_Neef2025}.
Our band construction recovers, in the observable-algebra language, the
encoded-subsystem universality established by Oreshkov
\cite{HQCsub_Oreshkov2009}: adiabatic loops can realize encoded-factor gates,
with ancillary and scalar phases treated as in that reference.
The present work determines the full holonomy of the canonical transport of
moving observable algebras (Theorems~\ref{thm:hol}--\ref{thm:WZ}) and identifies,
for sectors with $n_J,d_J>1$, band-resolved holonomies that act on the refinement
without defining automorphisms of the coarser algebra.

For at least two bands, Theorem~\ref{thm:WZ} realizes every joint holonomy
$h=\bigoplus_{J,a}h_{J,a}$ satisfying $h_{J,a}\in\mathrm U(d_J)$ and
$\prod_{J,a}\det h_{J,a}=1$ on the full ordered band flag. These matrices act
separately on the bands and can differ between copies of the same logical
subsystem. To act only on the logical factors, choose $h_{J,a}=W_J$ for every
band $a$ in sector $J$. Then $h=\bigoplus_J\Id_{n_J}\otimes W_J\in\Uofb(\A)$,
and the determinant condition becomes $\prod_J(\det W_J)^{n_J}=1$.
Thus, on every positive-dimensional stratum, closed loops of band structures
realize every logical unitary up to an overall phase, and hence every inner
automorphism of $\A$, without a genericity assumption (Appendix~\ref{sec:isospectral})
\cite{HQCuniv,HQCuniv_Lucarelli2002,HQCuniv_ZanardiLloyd2004,HQCuniv_Tanimura2005}.

Intrinsic transport and isospectral transport of the band refinement can differ
geometrically but have the same maximal attainable inner action on $\A$
(Appendix~\ref{sec:two}). The dynamical phase generated by $H_0\in\Ap$ acts trivially on
$\A$. Given access to the required isospectral path, the target inner automorphism
is therefore implemented in the adiabatic limit
\cite{Kato1950,AvronSeilerYaffe1987}.

\emph{Summary.}---We have established a geometric framework for processing
quantum information by parallel transport of the observable algebras that
define generalized quantum subsystems. Each stratum of fixed block structure carries a
unique unitarily invariant connection, recovering Berry--Wilczek--Zee transport
on bands. On every positive-dimensional stratum,
$\Hol=\Nrm\cap\mathrm{SU}(d)$: closed paths realize all inner logical gates
and equal-type sector permutations. On these strata, a compatible band
refinement attains the same inner action, although its transport need not
coincide with intrinsic algebra transport. Given a compatible gapped Hamiltonian
and access to the required isospectral paths of the refinement, every inner
logical gate can be implemented in the adiabatic limit.

\begin{acknowledgments}
P.Z. acknowledges partial support from the National Science Foundation under award
PHY-2310227. The author used Anthropic Fable~5.1 for manuscript drafting,
revision, and \LaTeX{} preparation, and GPT Astra~6 for technical review,
analytical and attribution checks, manuscript revision, and \LaTeX{} editing
and compilation. AI assistance with analytical checks and numerical code is
detailed in Appendix~\ref{sec:num}. The author specified the tasks, checked
the results, and reviewed and revised the AI-assisted material, and takes
responsibility for the work.
\end{acknowledgments}

\emph{Data availability.}---The numerical code, input seeds, and complete
results supporting this work are provided in the accompanying ancillary
package \texttt{hol-alg-sm-v3}, described in Appendix~\ref{sec:num}.

\clearpage
\onecolumngrid
\appendix
\numberwithin{equation}{section}
\numberwithin{table}{section}
\renewcommand{\theequation}{\thesection\arabic{equation}}
\renewcommand{\thetable}{\thesection\arabic{table}}

\noindent
These appendices supply the proofs and supporting calculations used in the
main text. We use its notation, the Hilbert--Schmidt inner product
$\langle X,Y\rangle=\Tr(X^\dagger Y)$, and $X_{\nLie}$ for orthogonal projection
onto $\nLie$. The groups $\Hol$ and $\Hol^0$ denote full holonomy and holonomy
from contractible loops, respectively.

\section{Algebraic splitting and leakage}
\label{sec:setup}

In the Wedderburn decomposition of Eq.~\eqref{eq:wedderburn}, write
$\HH_J=\mathbb C^{n_J}\otimes\mathbb C^{d_J}$ and let $\Pi_J$ project onto
$\HH_J$. The algebra $\A$ defines a \emph{generalized quantum subsystem}:
the center $\Z=\bigoplus_J\mathbb C\Pi_J$ carries the classical sector label,
while each $\HH_J$ has a tensor-product structure. For $Z=1$, $\A$ is a
factor and specifies an ordinary tensor-factor subsystem.

\emph{Normalizer and horizontal space.}
A normalizer permutes the minimal central projectors only among
equal-type sectors. If it fixes every sector, its automorphism of each matrix
factor is inner; removing that action leaves a unitary in the commutant. Thus
\begin{equation}\label{eq:normalizer}
 \Nrm_0=\left\{\bigoplus_J U_J\otimes W_J\right\},\qquad
 \pi_0(\Nrm)=\prod_\alpha S_{m_\alpha},\qquad
 \nLie=(\A+\Ap)\cap\uLie,
\end{equation}
where $U_J\in\mathrm U(n_J)$, $W_J\in\mathrm U(d_J)$, and $m_\alpha$ counts
sectors of the same type. The block products form a connected group; their
scalar redundancy gives $\dim\nLie=\sum_J(n_J^2+d_J^2-1)$.
All equal-type permutations are implemented by matching tensor-product bases.

For $X_J=\Pi_JX\Pi_J$, the conditional expectations are
\begin{equation}\label{eq:explicitP}
 \PA(X)=\bigoplus_J\frac{\Id_{n_J}}{n_J}\otimes\Tr_{n_J}X_J,\qquad
 \PAp(X)=\bigoplus_J\Tr_{d_J}X_J\otimes\frac{\Id_{d_J}}{d_J}.
\end{equation}
These are the Hilbert--Schmidt projections. A trace-preserving conditional
expectation $E$ onto $\A$ satisfies, for $c\in\A$,
\[
 \Tr(c^\dagger E(X))=\Tr E(c^\dagger X)=\Tr(c^\dagger X),
\]
so $E=\PA$. Applying the two partial traces in either order gives
$\PA\PAp=\PAp\PA=\PZ$. Consequently
\begin{equation}\label{eq:splitS}
 \PR=\id-\PA-\PAp+\PZ,\qquad
 \uLie=\nLie\oplus\mLie,\qquad \mLie=\Rsp\cap\uLie,
 \quad \Rsp=(\A+\Ap)^\perp.
\end{equation}
Writing $\Mat_0$ for traceless matrices, decomposition of each matrix factor
into its scalar and traceless parts yields
\begin{equation}\label{eq:Rexplicit}
 \Rsp=\bigoplus_J\Mat_0(n_J)\otimes\Mat_0(d_J)
       \ \oplus\ \bigoplus_{J\ne K}\LB(\HH_K,\HH_J).
\end{equation}
Thus $\dim\mLie=d^2-\sum_J(n_J^2+d_J^2-1)$ and every horizontal matrix is
traceless. Conjugation by any normalizer preserves $\A,\Ap$ and the
Hilbert--Schmidt inner product, hence both summands in \eqref{eq:splitS}.
The differential of the orbit map at the identity has kernel $\nLie$,
so it identifies $\mLie$ with the tangent
space of $\AlgSet(\btau)=\Ud/\Nrm$.
For completeness, representing an algebra by $\PA$ makes each type orbit
compact and connected. There are finitely many types at fixed $d$, so these
disjoint compact orbits are also open in their union and are its connected
components.

\emph{Leakage and squared commutators.}
For an algebra $\B$, define its canonical generators sectorwise by
$k^{(J)}_{ab}=d_J^{-1/2}\Id_{n_J}\otimes E^{(J)}_{ab}$, using the type of
$\B$. Then $\sum_\alpha\|k_\alpha\|_2^2=d$. Define
\begin{equation}\label{eq:leakage}
 K(\A,\B)=1-\frac1d\sum_\alpha\|\PA(k_\alpha)\|_2^2
 =\frac1{2d}\sum_{\alpha,\beta}\|[a'_\beta,k_\alpha]\|_2^2,
\end{equation}
where $a'^{(J)}_{ab}=n_J^{-1/2}E^{(J)}_{ab}\otimes\Id_{d_J}$ are the
canonical generators of $\Ap$. Each family uses its own algebra's
Wedderburn decomposition and tensor-product basis; their sector labels need
not coincide. This normalization is part of the definition.
To verify the identity, matrix-unit multiplication in \eqref{eq:explicitP}
gives $\sum_\beta a'_\beta Y a'^{\dagger}_\beta=\PA(Y)$, with the same
identity after exchanging each generator and its adjoint; at $Y=\Id$ both
sums equal $\Id$. Expanding the commutator norm therefore gives
\begin{equation}\label{eq:commid}
 \sum_\beta\|[a'_\beta,Y]\|_2^2
 =2\bigl(\|Y\|_2^2-\|\PA(Y)\|_2^2\bigr).
\end{equation}
Summing over $Y=k_\alpha$ proves \eqref{eq:leakage}. Orthogonal projection
also gives $0\le K\le1$ and $K=0$ precisely when $\B\subseteq\A$.

For $X\in\uLie$, use the canonical generators of $\A$ and put
$Q=\id-\PA$. Since $Qk_\alpha=0$,
\[
 K(\A,e^{tX}\A e^{-tX})
 =\frac{t^2}{d}\sum_\alpha\|Q[X,k_\alpha]\|_2^2+O(t^3).
\]
The bimodule identity gives $\PA[X,k]=[\PA X,k]$. Applying
\eqref{eq:commid} with $\A$ and $\Ap$ exchanged, and using the commuting
square, evaluates the sum as
$2(\|X\|_2^2-\|\PAp X\|_2^2-\|\PA X\|_2^2+\|\PZ X\|_2^2)
=2\|\PR X\|_2^2$. Hence
\begin{equation}\label{eq:Hessian}
 K(\A,e^{tX}\A e^{-tX})=\frac{2t^2}{d}\|\PR X\|_2^2+O(t^3).
\end{equation}
For $\B=U\A U^\dagger$, \eqref{eq:leakage} is an algebraic averaged
out-of-time-order commutator \cite{OTOC,AOTOC}. Equation~\eqref{eq:Hessian}
is the local operational interpretation of the horizontal directions.

\section{Uniqueness of the invariant connection}
\label{sec:unique}

For a reductive bundle $G\to G/K$, Wang's classification
\cite{Wang1958,KobayashiNomizu} writes every invariant connection as
$\omega_g(gX)=X_{\mathfrak k}+\Lambda(X_{\mathfrak m})$, where
$\Lambda:\mathfrak m\to\mathfrak k$ is $K$-equivariant.
Here $\Lambda$ would add a gauge rotation to a horizontal displacement.
Equivariance means $\Lambda(kXk^\dagger)=k\Lambda(X)k^\dagger$:
the prescription must respect changes of frame that preserve the algebra.
We can therefore test it using independent rotations of the tensor factors.

Here equivariance under $\Nrm_0$ alone forces $\Lambda=0$:
on an inter-sector block $\mLie_{JK}$, conjugation by
$\Id-2\Pi_J$ is $-1$, while it is $+1$ on $\nLie$.

On a nonzero within-sector block, the complexified representation is
$\Mat_0(n_J)\otimes\Mat_0(d_J)$, the tensor product of two nontrivial
adjoint representations. Each summand of the complexified $\nLie$
is trivial under at least one of these two factors. Averaging over that
factor's unitary rotations kills the traceless input but leaves the output
unchanged. Equivariance then forces that output to vanish. This is the
Schur-lemma argument in matrix form. These exhaust \eqref{eq:Rexplicit}, proving
uniqueness of Eq.~\eqref{eq:omega}.

The constructive lift in the main text follows directly. For
$\Omega=V^\dagger\dot V$ and $\dot h=-\Omega_{\nLie}h$, $h(0)=\Id$,
the solution lies in $\Nrm_0$, and
\begin{equation}\label{eq:lift}
 (Vh)^\dagger\frac{d}{dt}(Vh)=h^\dagger\Omega_{\mLie}h\in\mLie.
\end{equation}
If the algebra path closes, its holonomy relative to $U_0=V(0)$ is
$U_0^\dagger V(1)h(1)\in\Nrm$.

\emph{Relation to conditional-expectation transport.}
The reductive projection in expectation-orbit geometry
\cite{CEorbits,CEorbits_AS,CEorbits_AL} is precisely
$\PA+\PAp-\PZ$. Proposition~4.13 of Ref.~\cite{CEorbits_AS} states prior
uniqueness of the homogeneous reductive structure; the argument above explicitly
computes $\operatorname{Hom}_{\Nrm}(\mLie,\nLie)=0$ in finite dimension.
The induced transport agrees on $\A$ with the expectation propagator
$\dot G_t=[\dot E_t,E_t]G_t$ of Ref.~\cite{CEorbits_AL}, with $G_0=\id$.
For this comparison choose a horizontal lift with $\widetilde U(0)=\Id$
and define $E_t=\mathbb P_{\A_t}$, where
$\A_t=\widetilde U(t)\A\widetilde U(t)^\dagger$.
Then, for
$a_t=\widetilde U a\widetilde U^\dagger$, $a\in\A$,
$\PA[\widetilde U^\dagger\dot{\widetilde U},a]=0$ by trace cyclicity and
orthogonality to $\A$. Thus $E_t\dot a_t=0$. With
$L_t=\operatorname{ad}_{\dot{\widetilde U}\widetilde U^\dagger}$,
$\dot E_t=[L_t,E_t]$ implies
$[\dot E_t,E_t]a_t=L_ta_t-E_tL_ta_t=\dot a_t$.
The common initial value proves $G_t(a)=a_t$.
This equality is restricted to the algebra; on $\ker E_t$ the generators
are $E_tL_t$ and $L_t$, respectively. Endpoint conjugation on $\A$ has
kernel $\Uofb(\Ap)$, and the pair of actions on $\A,\Ap$ has kernel
$\Uofb(\Z)$.

\section{Intrinsic holonomy and worked examples}
\label{sec:curv}

\emph{(i)--(iii): curvature, determinant, and restricted holonomy.}
For horizontal left-invariant fields $\widetilde X(U)=UX$,
$[\widetilde X,\widetilde Y](U)=U[X,Y]$. Thus
$F=d\omega+\tfrac12[\omega\wedge\omega]$ gives
$F_U(UX,UY)=-[X,Y]_{\nLie}$. These curvature values are the same at every
point of the holonomy subbundle, so Ambrose--Singer \cite{AmbroseSinger1953}
gives
\begin{equation}\label{eq:holalg}
 \holLie^0=\spann_{\mathbb R}\{[X,Y]_{\nLie}:X,Y\in\mLie\}.
\end{equation}
In particular this span is a Lie algebra. Horizontality also gives
$\Tr(\widetilde U^\dagger\dot{\widetilde U})=0$, so the lift preserves
its determinant and every holonomy has determinant one. For the positively
oriented coordinate square with lift
$V(s,t)=e^{s\varepsilon X}e^{t\varepsilon Y}$, first traversed along $X$,
the sign in $\dot h=-\omega(\dot V)h$ gives
\begin{equation}\label{eq:plaquette}
 \hol=\exp\!\left(\varepsilon^2[X,Y]_{\nLie}\right)+O(\varepsilon^3).
\end{equation}

\emph{(iv): flatness.}
If $F=0$, then $[\mLie,\mLie]\subseteq\mLie$; reductivity makes
$\mLie$ an ideal of $\uLie$. Since $\mLie$ is traceless and $\su(d)$ is
simple for $d\ge2$, either $\mLie=0$ or $\mLie=\su(d)$. The latter would
force $\nLie=i\mathbb R\Id$, hence $\A+\Ap=\mathbb C\Id$, which is
impossible for $d\ge2$. Thus flatness is equivalent to $\mLie=0$, a point
stratum. The case $d=1$ is a point as well.

\emph{(v): outer components.}
The exact sequence of $\Nrm\to\Ud\to\AlgSet(\btau)$ contains
\begin{equation}\label{eq:homotopy}
 \pi_1(\Nrm)\longrightarrow\mathbb Z
 \longrightarrow\pi_1(\AlgSet(\btau))
 \longrightarrow\pi_0(\Nrm)\longrightarrow1.
\end{equation}
The last map records the component of a lift's endpoint, hence of its
holonomy. It is onto. By \eqref{eq:normalizer}, its kernel corresponds to
inner actions, while every nonidentity permutation moves a minimal central
projector and is outer.

\emph{(vi): full holonomy.}
The aim is to generate every allowed gauge rotation using horizontal
controls. Their commutators supply two kinds of motion: rotations within
each tensor factor and relative phases between sectors. We first show that
curvature generates all these directions, then pass from infinitesimal
generators to finite pulse sequences.
Assume $\mLie\ne0$ and put $\mathfrak h=\holLie^0$. The span
\eqref{eq:holalg} is $\Ad(\Nrm)$-invariant and traceless, hence is an
ideal in
\begin{equation}\label{eq:nsu}
 \nLie\cap\su(d)=
 \bigoplus_J\bigl(\su(n_J)\oplus\su(d_J)\bigr)\oplus\mathfrak t,
 \qquad
 \mathfrak t=\left\{\sum_Jic_J\Pi_J:\sum_Jc_Jn_Jd_J=0\right\}.
\end{equation}
The space $\mathfrak t$ consists of sector phases whose weighted sum is
zero, removing the overall phase forbidden by determinant preservation.
An ideal contains each simple factor on which it has nonzero projection.
When $n_J,d_J\ge2$, horizontal matrices $X=ia\otimes p$,
$Y=ib\otimes p$, with $a,b,p$ traceless Hermitian, give
\[
 [X,Y]_{\nLie}=-\frac{\Tr p^2}{d_J}[a,b]\otimes\Id_{d_J}.
\]
These span $\su(n_J)$; exchanging the factors supplies $\su(d_J)$.
If there are at least two sectors, inter-sector blocks supply the same
factors even when a tensor factor has dimension one. For maps
$x,y:\HH_K\to\HH_J$, let $X=x-x^\dagger$, $Y=y-y^\dagger$. Then
\begin{equation}\label{eq:crossblocks}
 [X,Y]_{JJ}=yx^\dagger-xy^\dagger,\qquad
 [X,Y]_{KK}=y^\dagger x-x^\dagger y.
\end{equation}
Taking $x=|a,p\rangle\langle b,q|$ and
$y=|a',p\rangle\langle b,q|$, with $a\perp a'$ and all vectors normalized,
gives a nonzero $\su(n_J)$ projection in the first block and zero in the
second. Here $a,a'$ are multiplicity vectors and $p$ is a logical vector
in sector $J$; $b,q$ are the respective vectors in sector $K$.
Interchanging multiplicity and logical indices gives $\su(d_J)$.
Finally, choosing $y=ix$ gives opposite block traces and central projection
\[
 2i\|x\|_2^2\left(\frac{\Pi_J}{n_Jd_J}
                         -\frac{\Pi_K}{n_Kd_K}\right).
\]
Subtracting the simple-factor parts already obtained leaves all of
$\mathfrak t$. For one sector, $\mathfrak t=0$ and $\mLie\ne0$ forces
both factors to be nontrivial. Therefore
$\mathfrak h=\nLie\cap\su(d)$ in every positive-dimensional case.

The Lie algebra generated by $\mLie$ contains
$[X,Y]_{\nLie}=[X,Y]-[X,Y]_{\mLie}$, hence contains $\mathfrak h$.
Since $\mLie\oplus\mathfrak h=\su(d)$, the connected subgroup generated
by $\exp\mLie$ is all of $\mathrm{SU}(d)$. Products of these exponentials
give exact reachability by the Lie-group controllability theorem
\cite{DAlessandro2001}: generators of the Lie algebra generate the compact
connected group by finite products, with real pulse durations. Both signs
are available here because $X\in\mLie$ implies $-X\in\mLie$.
These products are endpoints of concatenated horizontal paths; smooth reparametrization
at the joins gives smooth paths with the same endpoints. Thus every element
of $\Nrm\cap\mathrm{SU}(d)$ is reached over the initial algebra.
The determinant constraint proves
\begin{equation}\label{eq:HolAll}
 \Hol=\Nrm\cap\mathrm{SU}(d),\qquad
 \Hol^0=(\Nrm\cap\mathrm{SU}(d))_0.
\end{equation}

\emph{Qubit and equal-type exchanges.}
For the qubit lift used in Eq.~\eqref{eq:qubithol}, direct differentiation gives
\[
 V^\dagger\dot V=-\frac i2\cos\theta\,\dot\varphi\,\sigma_z
 +\frac i2\sin\theta\,\dot\varphi\,\sigma_x
 -\frac i2\dot\theta\,\sigma_y.
\]
The first term is the gauge part. For azimuthal winding $w$,
$V(0)^\dagger V(1)=(-1)^w\Id$; multiplication by
$h(1)=\exp[\frac i2\sigma_z\oint\cos\theta\,d\varphi]$ gives
$\hol=e^{-i\Omega_s\sigma_z/2}$ with
$\Omega_s=2\pi w-\oint\cos\theta\,d\varphi$. The horizontal path
$e^{-i\pi t\sigma_x/2}$ gives the other component, $-i\sigma_x$.

More generally, let $L:\HH_K\to\HH_J$ identify corresponding tensor
bases of two equal-type sectors, extended by zero. For physical duration $T$,
\begin{equation}\label{eq:exchange}
 K=\frac{\hbar\pi}{2T}(L+L^\dagger),\qquad
 U(t)=e^{-i\pi t(L+L^\dagger)/(2T)},\qquad
 U(T)=\Id-\Pi_J-\Pi_K-i(L+L^\dagger).
\end{equation}
The velocity $U^\dagger\dot U$ is off diagonal between sectors, hence
horizontal. The endpoint exchanges their algebras and has determinant one.
This proves the sector-exchange construction: the physical pulse is dynamical,
and its holonomy interpretation refers to the closed algebra path.

\emph{Stabilizer codes.}
Let $S=\langle g_1,\ldots,g_r\rangle$ be a Pauli stabilizer on $n$ qubits:
the $r$ generators are independent and commute, $-\Id\notin S$, and
$1\le r\le n$ \cite{Gottesman1997}. A binary syndrome
$x=(x_1,\ldots,x_r)$ labels the joint eigenspace $\HH_x$ in which
$g_j$ has eigenvalue $(-1)^{x_j}$. There are $q=2^r$ such sectors,
each of dimension $m=2^{n-r}$ and carrying $n-r$ logical qubits.
The all-$+1$ sector $\HH_0$ is the code space.

Writing $\Pi_x$ for the syndrome projectors, the algebra
$\A_S=\spann_{\mathbb C}S=\bigoplus_x\mathbb C\Pi_x$ describes
syndrome observables, while
$\A_S'=\bigoplus_x\LB(\HH_x)$ contains all observables that preserve
each syndrome sector, including its logical observables.
Since $\A_S$ is commutative, the common gauge projection for these two
algebras reduces to $X\mapsto\sum_x\Pi_xX\Pi_x$.
Thus block-diagonal generators change frames within sectors, while
horizontal generators couple different syndrome sectors. This explains
why the syndrome algebra and its commutant have the same transport.

All sectors have the same dimension, so their common normalizer allows
independent unitaries within sectors and arbitrary sector permutations:
$N(\A_S)=\mathrm U(m)^q\rtimes S_q$.
As $q\ge2$, \eqref{eq:HolAll} gives Eq.~\eqref{eq:stabilizer-hol}.
Contractible algebra loops return each syndrome sector to itself and realize
\[
 \Hol^0=S(\mathrm U(m)^q)
 =\left\{\bigoplus_xV_x:\ V_x\in\mathrm U(m),\quad
                         \prod_x\det V_x=1\right\}.
\]
This group is connected: the allowed block determinants form a
$(q-1)$-torus, and at fixed determinants the remaining freedom is the
connected group $\mathrm{SU}(m)^q$. The determinant constraint couples
only the block phases; the logical rotations are otherwise independent.
These unitaries commute with the syndrome observables in $\A_S$ but
can act nontrivially on encoded states and logical observables in $\A_S'$.

The other components exchange syndrome projectors. Every permutation
has a determinant-one representative after an overall phase adjustment,
so $\Hol/\Hol^0\cong S_q$. Equation~\eqref{eq:exchange} gives an explicit
horizontal pulse exchanging any pair of sectors through couplings between
corresponding basis states. Such an exchange closes the algebra path even
though the individual syndrome projectors do not return to themselves.

\section{Tensor factors and winding classes}
\label{sec:factor}

\emph{Curvature and the two-qubit exception.}
For $\A=\Id_{d_1}\otimes\Mat(d_2)$, $d_1,d_2\ge2$, formula
\eqref{eq:explicitP} identifies horizontal controls as the interactions
with both partial traces zero. For traceless Hermitian $a,c$ and $b,e$,
the tensor commutator identity gives
\begin{align}
 [ia\otimes b,ic\otimes e]_{\nLie}
 &=-\frac{\Tr(be)}{d_2}[a,c]\otimes\Id
   -\frac{\Tr(ac)}{d_1}\Id\otimes[b,e],\label{eq:curvTPS}\\
 [ia\otimes b,ic\otimes e]_{\mLie}
 &=-\tfrac12\bigl([a,c]\otimes\{b,e\}_0
                         +\{a,c\}_0\otimes[b,e]\bigr),\label{eq:horizontalTPS}
\end{align}
where the subscript $0$ removes the scalar part. The curvature is the
negative of \eqref{eq:curvTPS}. For two qubits all such anticommutators
are scalar, so $[\mLie,\mLie]\subseteq\nLie$. In particular,
$[i\sigma_x\otimes\sigma_x,i\sigma_x\otimes\sigma_y]
=-2i\Id\otimes\sigma_z$, giving the small-loop gate in the main text by
\eqref{eq:plaquette}.
If $d_2\ge3$, choose
$b=\operatorname{diag}(1,-1,0,\ldots)$,
$e=\operatorname{diag}(1,1,-2,0,\ldots)$ and embedded Pauli matrices
$a=\sigma_x$, $c=\sigma_y$. Both trace overlaps vanish, but the full
commutator is $-[a,c]\otimes b\ne0$ and is horizontal. Exchanging factors
covers $d_1\ge3$. Thus only two qubits have the stated symmetric
splitting \cite{Helgason,Besse1987}, and only there do vanishing overlaps
always force the full commutator to vanish.

\emph{Sector permutations and winding classes.}
The fundamental group $\pi_1$ classifies algebra loops up to continuous
deformation, whereas $\pi_0$ records disconnected components of their lift
endpoints. Sector permutations account for some loop classes; determinant
winding detects the remaining classes even when every sector returns to itself.
Let $k=\gcd\{n_J,d_J\}_J$ and
$\mathsf P=\pi_0(\Nrm)=\prod_\alpha S_{m_\alpha}$.
The map $\prod_J[\mathrm U(n_J)\times\mathrm U(d_J)]\to\Nrm_0$
has connected kernel
$\{(\lambda_J\Id,\lambda_J^{-1}\Id)_J\}$, so it is onto on
fundamental groups. Since
$\det(\bigoplus_JU_J\otimes W_J)
=\prod_J(\det U_J)^{d_J}(\det W_J)^{n_J}$,
the possible determinant windings in $\Nrm_0$ form $k\mathbb Z$.
Indeed, winding $\det U_J$ once winds the full determinant $d_J$ times,
and winding $\det W_J$ once contributes $n_J$ turns. Combining these
integer windings gives precisely the multiples of their greatest common
divisor. Thus winding is distinguished only modulo $k$ after quotienting
by normalizer loops.
Equation~\eqref{eq:homotopy} therefore gives
\begin{equation}\label{eq:centralextension}
 1\longrightarrow\mathbb Z_k\longrightarrow
 \pi_1(\AlgSet(\btau))\longrightarrow\mathsf P\longrightarrow1.
\end{equation}
Equation~\eqref{eq:centralextension} identifies $\pi_1(\AlgSet(\btau))$
as an extension of the sector-permutation group $\mathsf P$ by the cyclic
winding group $\mathbb Z_k$. The map to $\mathsf P$ remembers only the
final permutation; its kernel consists of the $k$ loop classes with no
sector permutation. On band strata $k=1$, so only sector permutations
remain.

\emph{Tensor-factor holonomy.}
For a single sector with $d_1,d_2\ge2$, set $d=d_1d_2$,
$k=\gcd(d_1,d_2)$ and $z=e^{2\pi i/d}\Id$.
Equation~\eqref{eq:nsu} gives
$\Hol^0=\mathrm{SU}(d_1)\otimes\mathrm{SU}(d_2)$.
By \eqref{eq:HolAll}, every holonomy is a tensor product
$U_1\otimes U_2$ with total determinant one. Removing a scalar phase
from each factor makes it special unitary, so the product has the form
$e^{i\phi}h_0$, with $h_0\in\Hol^0$. Its determinant is $e^{id\phi}$;
hence $e^{i\phi}h_0=z^jh_0$ for some integer $j$.

To determine which powers of $z$ already belong to $\Hol^0$, note that
a tensor product is scalar only if both factors are scalar. Their determinant-one conditions
on those factors therefore give
\[
 z^j\in\Hol^0
 \quad\Longleftrightarrow\quad
 j\equiv ad_2+bd_1\pmod d\quad\text{for some }a,b\in\mathbb Z
 \quad\Longleftrightarrow\quad k\mid j,
\]
where the last equivalence is Bezout's identity. There are thus exactly
$k$ distinct cosets, each connected because $\Hol^0$ is connected.
There is no sector permutation in this case ($\mathsf P=1$), so
\eqref{eq:centralextension} also gives $\pi_1(\AlgSet(\btau))\cong\mathbb Z_k$.
Together these facts prove Theorem~\ref{thm:factor}:
\begin{equation}\label{eq:Hol22S}
 \Hol=\bigsqcup_{j=0}^{k-1}z^j\Hol^0,
 \qquad \Hol/\Hol^0\cong\pi_1(\AlgSet(\btau))\cong\mathbb Z_k.
\end{equation}
Overall phases do not change the algebra, so the stratum is also
$\mathrm{SU}(d)/\Hol$. Since $\mathrm{SU}(d)$ is simply connected,
two algebra loops are homotopic precisely when their lifted endpoints
lie in the same holonomy component. This identifies the loop classes
with the component labels in \eqref{eq:Hol22S}.
The factors $z^j$ act trivially by conjugation, so the component label
adds no observable automorphism; the factor in $\Hol^0$ can still depend
on the loop's shape.

For two qubits, the antiunitary
$C=-(\sigma_y\otimes\sigma_y)\widehat K$, with $\widehat K$ complex
conjugation, has $C^2=\Id$ and commutes with
$\mathrm{SU}(2)\otimes\mathrm{SU}(2)$. In a $C$-real (magic) basis
\cite{MagicBasis}, this connected six-dimensional group is $\mathrm{SO}(4)$.
Thus $\Hol=\mathrm{SO}(4)\sqcup i\,\mathrm{SO}(4)$ and
$\AlgSet(\{(2,2)\})\cong(\mathrm{SU}(4)/\mathrm{SO}(4))/\mathbb Z_2$,
as stated in the main text.

\emph{Controlled-phase generator.}
For rank-one projectors $P,Q$, the loop
$\gamma(t)=e^{2\pi itP\otimes Q}\A e^{-2\pi itP\otimes Q}$ closes
at $t=1$. Set
\begin{equation}\label{eq:cphase}
 A_0=2\pi i\left(P-\frac{\Id}{d_1}\right)
                  \otimes\left(Q-\frac{\Id}{d_2}\right),\qquad
 e^{tA_0}=e^{2\pi itP\otimes Q}g(t),\quad
 g(t)=e^{2\pi it/d}e^{-2\pi itP/d_2}\otimes e^{-2\pi itQ/d_1}.
\end{equation}
The commuting factors verify this identity. Since $A_0\in\mLie$ and
$g(t)\in\Nrm_0$, $e^{tA_0}$ is the horizontal lift; $A_0\ne0$ also
shows the algebra path is nonconstant. At $t=1$, its endpoint is
\[
 g(1)=z^{-1}\left(
 e^{-2\pi i(P-\Id/d_1)/d_2}\otimes
 e^{-2\pi i(Q-\Id/d_2)/d_1}\right).
\]
Both local generators are traceless, so the parenthesized factor belongs
to $\Hol^0$. Thus $g(1)\in z^{-1}\Hol^0$: its component label is
$-1\bmod k$, a generator, and $m$ repetitions lie in $\Hol^0$ exactly
when $k\mid m$.

\section{Recovery of Berry--Wilczek--Zee transport}
\label{sec:band}

On a band stratum, $n_J=1$ or $d_J=1$ in every sector, so
$\A_J+\Ap_J=\LB(\HH_J)$. Thus $\nLie$ is block diagonal and
$\mLie$ is block off diagonal. For
$|\psi_a\rangle=\widetilde U|a\rangle$, horizontality is exactly
$\langle\psi_a|\dot\psi_b\rangle=0$ within each band, the WZ condition
\cite{WZ}. Eigenvalue-labelled bands use the ordered flag
$\mathcal F_{\mathbf d}=\Ud/\prod_\mu\mathrm U(d_\mu)$
\cite{Spectral1993}; forgetting labels among equal-type bands gives the algebra
stratum. This covering preserves the connection locally, while the
ordered flag excludes band permutations. Projection from the ordered flag
to a single band's Grassmannian pulls back the universal Stiefel connection
to that band's WZ connection \cite{NarasimhanRamanan61}. Pullback along a Hamiltonian family
then gives the joint Berry--WZ transport.

\section{Isospectral holonomy and logical universality}
\label{sec:isospectral}

\emph{Ordered-band holonomy from Theorem~\ref{thm:hol}.}
Use the compatible $H_0=\bigoplus_JH_J\otimes\Id_{d_J}$ of Eq.~\eqref{eq:family}, with each $H_J$ nondegenerate, disjoint sector spectra, and at
least two bands. Its band refinement is
$\At=\{H_0\}''\vee\A=\bigoplus_{\mu=(J,a)}\Mat(d_J)$, with
$d_\mu=d_J$. The ordered flag is $\Ud/B$, where
$B=\prod_\mu\mathrm U(d_\mu)=N(\At)_0$ fixes each labelled band.
By Appendix~\ref{sec:band}, its connection has the same horizontal lifts as
the canonical connection for $\At$. Theorem~\ref{thm:hol}(ii),
applied to $\At$, therefore gives determinant one for the
joint flag holonomy.

With at least two bands, the orbit of $\At$ is positive-dimensional.
Theorem~\ref{thm:hol}(vi) realizes every element of $N(\At)\cap\mathrm{SU}(d)$
as the endpoint of a horizontal path starting at $\Id$.
Such a path closes on the ordered flag precisely when its endpoint
lies in $B$. Restricting the endpoints therefore gives
\[
 \Hol_{\mathrm{flag}}=B\cap\mathrm{SU}(d)
 =S(B):=\left\{(g_\mu)\in B:\prod_\mu\det g_\mu=1\right\}.
\]
Every specified family is a pullback from the flag and hence
has holonomy contained in $S(B)$. The containing group has dimension
$\sum_J n_Jd_J^2-1$. This proves Theorem~\ref{thm:WZ} without
an assumption of genericity.

In an eigenframe, write $\theta=U^\dagger dU$ and
$A^{(\mu)}=\theta_{\mu\mu}$. Theorem~\ref{thm:hol}(i) gives the block curvature
$F^{(\mu)}=-\sum_{\nu\ne\mu}\theta_{\mu\nu}\wedge\theta_{\nu\mu}$.
The traceless curvature and determinant-one holonomy of Theorem~\ref{thm:hol}
thus read, in band notation,
\begin{equation}\label{eq:sumruleS}
 \sum_\mu\Tr F^{(\mu)}=0,\qquad \prod_\mu\det h^{(\mu)}=1.
\end{equation}
These identities remain valid under pullback; the determinant identity
holds for every loop in the control manifold, including noncontractible
ones. On the full ordered flag, each band can
realize $\mathrm{SU}(d_\mu)$, and the band determinants fill a torus of
dimension $\sum_J n_J-1$; individual determinants need not be one.
This is the collective
determinant refinement of the group in Ref.~\cite{VQS}.

\emph{Specified families and the conjecture.}
A pullback need not attain the flag's holonomy: a constant family already
has trivial holonomy. Attainment is open in the $C^2$ topology, because
Ambrose--Singer witnesses the full Lie algebra by finitely many curvature
values transported to a common base point and their brackets. Along fixed
paths these depend continuously on $U$; a nonzero full-rank minor persists
under a small perturbation. Density is the unproved part:
\begin{conjecture}[Generic attainment]\label{conj:generic}
Fix a compatible $H_0$ with at least two bands. For compact connected $M$
with $\dim M\ge2$, the maps $U\in C^\infty(M,\Ud)$ for which the family
$H=UH_0U^\dagger$ satisfies
$\Hol^0_{\mathrm{WZ}}=S(\prod_\mu\mathrm U(d_\mu))$ form a dense subset
of $C^\infty(M,\Ud)$ in the $C^2$ topology.
\end{conjecture}
Appendix~\ref{sec:num} gives finite numerical evidence only.

\emph{Logical operations and adiabatic implementation.}
The logical unitary $\bigoplus_J\Id_{n_J}\otimes W_J$ embeds with the
same $W_J$ on every band of sector $J$. Its determinant is
$\prod_J(\det W_J)^{n_J}$, so
\begin{equation}\label{eq:logicalintersection}
 S(B)\cap\Uofb(\A)=\Uofb(\A)\cap\mathrm{SU}(d)
 =:\mathcal{SU}(\A).
\end{equation}
A scalar phase enforces determinant one without changing conjugation.
Thus $\mathcal{SU}(\A)$ induces every inner automorphism.
For $V\in\mathcal{SU}(\A)$, full flag holonomy supplies a loop with
horizontal lift $\widetilde U(0)=\Id$, $\widetilde U(1)=V$.
Then $H(t)=\widetilde U(t)H_0\widetilde U(t)^\dagger$ is isospectral
and closes because $[V,H_0]=0$. Reparametrizing the loop to be constant
near its endpoints makes the control smoothly periodic. Given access
to this path, the adiabatic theorem \cite{Kato1950,AvronSeilerYaffe1987}
implements $V$ together with a dynamical phase in
$\Uofb(\{H_0\}'')\subseteq\Uofb(\Ap)$, which acts trivially on $\A$.
This recovers encoded-subsystem universality \cite{HQCsub_Oreshkov2009}; exactness
concerns geometric reachability, with physical implementation in the
adiabatic limit.

A general band tuple $(g_{J,a})$ normalizes the coarser $\A$ only when
its conjugation actions agree for all $a$ in each sector, equivalently
$g_{J,a}=e^{i\phi_{J,a}}W_J$. If $n_J,d_J>1$, special-unitary rotations
can differ nontrivially between copies while obeying the determinant
constraint. Such holonomies act on $\At$ but do not normalize $\A$,
as asserted in the main text.

\section{Intrinsic transport versus band-refinement transport}
\label{sec:two}

For $\A=\Id_2\otimes\Mat(2)$ and
$H_0=H_1\otimes\Id_2$ with $H_1$ nondegenerate, the intrinsic stratum
has dimension $9$ and full holonomy
$\mathrm{SO}(4)\sqcup i\,\mathrm{SO}(4)$ (in the magic basis).
The ordered flag of the two bands is $\mathrm{Gr}(2,\mathbb C^4)$,
of dimension $8$, with connected holonomy $S(\mathrm U(2)\times\mathrm U(2))$.
The holonomy groups have dimensions $6$ and $7$, respectively. A band
tuple $(g_1,g_2)$ normalizes $\A$ exactly when $g_1$ and $g_2$ differ
by a scalar, so generic flag holonomies do not.

Nevertheless both holonomy groups intersect $\Uofb(\A)$ in
$\mathcal{SU}(\A)=\{\Id_2\otimes W:(\det W)^2=1\}$:
this follows from \eqref{eq:HolAll} for intrinsic transport and
\eqref{eq:logicalintersection} for the flag. Both therefore induce all
inner automorphisms of $\A$. The same two intersection arguments apply
to every positive-dimensional algebra stratum and its compatible band
refinement. Equality of this attainable action does not identify the
two connections or their paths.

\section{Reproducible numerical checks}
\label{sec:num}

The ancillary package \texttt{hol-alg-sm-v3} supplies six executable
programs, complete results, seeds, enforced tolerances, software versions,
and source hashes. All six programs pass. Its coverage includes all
$70$ algebra types with $d=2$--$6$, twelve tensor factorizations, direct
connection-ODE transport, six stabilizer patterns, and five equal-type
exchanges. Structural ranks use threshold $10^{-9}$; algebraic-identity residuals
are below $10^{-12}$ and the relative Hessian error at $t=10^{-4}$ is
below $10^{-6}$. The README gives the per-program cases and thresholds.
These checks support the analytic calculations and enter no proof.

For sampled attainment, use Gaussian anti-Hermitian matrices $G_j$ and set
\[
 U(x,y)=\exp\!\left(\sum_{j=1}^{5}c_j(x,y)G_j\right),\qquad
 (c_1,\ldots,c_5)=(1,\sin x,\cos x,\sin y,\cos y).
\]
Seeds $101,202,303$ give three sampled unitary families. Each is tested with
both band patterns $(1,1,1,1)$ and $(2,2)$, giving six cases; the generators
and sampling points are shared between the two patterns for each seed.
Using exponential Fr\'echet derivatives, we evaluate curvature at twelve
points in $[-1,1]^2$ and integrate $\dot h=-A(\dot\lambda)h$ along
straight paths from $(0.2,0.3)$, with relative tolerance $10^{-10}$ and
absolute tolerance $5\times10^{-12}$. The matrices compared are
$h^\dagger F_{12}h$, all expressed at that common base. This transport makes
their comparison independent of unrelated local frame choices.
Their real spans have ranks $3$ and $7$ at threshold $10^{-7}$, matching
the predicted dimensions of $S(B)$ and counting independent infinitesimal
holonomy generators. Smallest retained singular
values exceed $0.48$ and $0.09$, and discarded values are below
$6\times10^{-11}$. Repeating seed $303$ at relative tolerance $10^{-12}$
preserves both ranks. Negative controls confirm gauge covariance and
show why pooling differently based loop logarithms, or arbitrary finite
principal logarithms, can give false dimensions. These finite samples
do not prove Conjecture~\ref{conj:generic}.

GPT Astra~6 assisted with analytical checks. GPT Astra~6 and
Anthropic Fable~5.1 assisted with preparation and revision of the
numerical-check code. The author directed these tasks, checked the results,
and reviewed and revised the AI-assisted material. Numerical verification
used the analytical identities, independent connection-ODE comparisons,
tightened-tolerance runs, and negative controls described above.
Full execution details are retained in the ancillary package.

\section{Scope}
\label{sec:scope}

The intrinsic results concern the principal connection on
$\Ud\to\Ud/\Nrm$, not Levi--Civita holonomy on the tangent bundle.
Adiabatic implementation uses a compatible gapped Hamiltonian and
accessible paths of its band refinement; it does not identify arbitrary
intrinsic algebra paths with adiabatic evolution. Full-flag holonomy
and designed logical reachability are proved. Attainment by generic
fixed families remains conjectural, and a constrained family can have
a proper subgroup.

\clearpage
\twocolumngrid

\end{document}